\documentclass[epj]{svjour}
\usepackage{graphicx}  
\usepackage{comment}
\usepackage{amsmath} 
\usepackage{xcolor} 
\usepackage{subfigure}
\usepackage{hyperref}  
\usepackage{cite}

\usepackage{booktabs} 

\begin{document}
\title{Neutron matter at the interface(s)} 
\subtitle{Static response and effective mass}
\author{Mateusz Buraczynski, Nawar Ismail, Alexandros Gezerlis}

\institute{Department of Physics, University of Guelph, Guelph, Ontario N1G 2W1, Canada}
\date{Received: date / Revised version: date}
%

\titlerunning{Effective Mass in Neutron Matter}

\abstract{
Neutron matter is interesting both as an extension of terrestrial nuclear physics and due to its significance for the study
of neutron stars. In this work, after some introductory comments on nuclear forces, nuclear \textit{ab initio} theory,
and nuclear phenomenology, we employ two techniques, Quantum Monte Carlo (QMC) and Energy Density Functionals, to practically handle 
an extended system composed of strongly interacting neutrons. We start by summarizing work
on the static response of neutron matter, which considers the impact of external influences on the time-independent system. 
We then proceed to discuss new results of the energy of quasiparticle excitations in neutron matter, including 
QMC calculations with chiral or phenomenological nucleon-nucleon interactions.
As part of this study, we carefully study the approach 
of our finite-number computations toward the infinite-system limit.
} 
\maketitle
\section{Introduction}\label{intro}

\subsection{Conceptual questions}

Many, if not most, working scientists are driven by an urge to get as close as possible to what
Bernard Williams termed the ``absolute conception of reality''\cite{Williams:1978}. 
While any human endeavor involves theories and interpretations,
the term introduced by Williams 
refers to what is there ``anyway'', i.e., what would be there even if there wasn't anybody there to investigate it.
To restate the same point, work in science has historically been motivated by a wish to ``get it right''. 
What this means in practice is harder to resolve; for example, in the study of the foundations of quantum mechanics such an outlook 
turns into a \textit{realist} approach, in contradistinction to an \textit{instrumentalist} one~\cite{Weinberg:2015}, with insufficient guidance
from experiment to decide in favor of one or the other. 
The fact that the interpretation of such a wildly successful theory as quantum mechanics is still an open research
question is cause for humility; it should therefore come as no surprise that applications of quantum mechanics which involve
few- to many-particle systems also raise a host of conceptual difficulties. 

This brings to mind an 
oft-quoted example of scientists' (supposed) hubris, Paul Dirac's 1929 phrase 
``\textit{The underlying physical laws necessary for the mathematical theory of a large part of physics and the whole of chemistry are thus completely known}''\cite{Dirac:1929}. What is less often mentioned is that Dirac went on to explicitly 
state that ``\textit{the difficulty is only that the exact application of these laws leads to equations much too complicated to be soluble}''. This, though indubitably an understatement, is an unmistakable admission that having access to an overarching theory does not imply 
a clear explanation of the wealth of observed (or not-yet-observed) states of matter. 
A more recent example along the same lines involves 
the theory of Quantum Chromodynamics (QCD): while we have not encountered the limits of QCD, this does
not mean that we are able to solve any single problem involving quark and gluon degrees of freedom.
To take a specific case: the physics of two or three strongly interacting neutrons, which will turn out to be very significant 
in the rest of this paper, is still not pinned down at the level of QCD.
Accomplishing this task is a goal of the \textit{reductionist} approach, which starts from a fundamental theory and the 
corresponding degrees of freedom.

Even if that goal \textit{can} be reached, it is no small undertaking. For now, it is safe to acknowledge that, 
even when an underlying fundamental theory is known, this does not necessarily imply predictive power to explain a given
phenomenon, practically speaking.
The question then arises how to make progress on the question of nuclear interactions: 
one, historically significant, approach
is to come up with nuclear-force models which do a great job of capturing neutron-proton scattering properties without
worrying about any possibly underlying level. A more recent
tack has been to employ Effective Field Theory (EFT), of the chiral or pionless variety, in an attempt to systematically
capture the known low-energy physics, while at the same time stating the limits of the applicability of the theory ahead of time;
this is conceptually pleasing, given its attempt to build bridges with the symmetries of the underlying theory, but of necessity
limited to low-energies. The passage from QCD to chiral EFT is the first of the interfaces mentioned in our title.
It's worth highlighting that, despite first appearances, the EFT philosophy is not necessarily in opposition to 
a reductionist outlook; after all, Steven Weinberg, the originator of the idea of Effective Field Theory, 
is one of the most vocal proponents of reductionism in living memory. In the future, matching QCD and an EFT for nuclear forces 
can benefit both approaches.

The present work addresses neutron matter, which is relevant to \textit{dense} objects such as 
neutron stars; we are therefore forced to consider how to extrapolate from terrestrially known nuclear
physics to more exotic settings. Today, this is not a mere toy problem: we are living in the era of multimessenger astronomy,
where the effect of a single neutron-star merger may foreseeably be measured using electromagnetic, gravitational-wave, and neutrino
signals. A few comments on the use of different nuclear interactions when studying compact stars may be in order. 
Phenomenological nuclear forces have the advantage that they can describe two-nucleon
physics up to very high energies; unfortunately, they also involve an arbitrariness in the three-nucleon
interaction, which can have a dramatic impact on the equation-of-state (EOS) of neutron-star matter. On the other hand, 
chiral EFT interactions have the conceptual advantage of following from a power expansion which provides guidance on
which terms (e.g., three-body forces) to include; as mentioned above, they have the limitation of being, by construction,
a low-energy effective theory, which should therefore not be used at densities of, say, 5 times that of nuclei on earth.
In short, each approach has its advantages, which is why both are still employed in practice.

Of course, even if one has made a choice in favor of a given nuclear-force approach, describing neutron matter 
needs another level (the second of the interfaces in our title), namely a \textit{quantum many-body approach}. 
To be explicit, this is the interface between few- and many-body physics.
Here, too, one can distinguish between two large classes: first, phenomenological approaches like energy-density
functionals have the advantage of a quasi-universal reach, being able to describe heavy to mid-mass nuclei very well;
one disadvantage relates to their predictive power: can the parameters that have been fit to experimental data be trusted
in regions where no experiment has taken place? The second large class includes \textit{ab initio} many-body theory; 
one should immediately note that the ``first'' principles spoken of in this context are \textit{not} the ones involving
the most fundamental degrees of freedom possible. Instead, \textit{ab initio} in the context of many-body theory
implies taking the degrees of freedom and the interactions between particles as given, and then trying to describe
a many-particle system without any free parameters. Once again, each approach has things to recommend it.

In the present paper, we report on our progress in working  
on a third interface, that between \textit{ab initio} many-body theory and phenomenological
many-body theory. As it so happens, we do this in two separate contexts: a) in section~\ref{sec:response}
we summarize our earlier work on the static response of neutron matter, where Density-Functional theory was
matched onto Quantum Monte Carlo, and b) in the rest of the paper we report on original derivations and computations
of the effective mass in neutron matter; this is a first step toward matching Landau's Fermi liquid theory of quasiparticles
to Quantum Monte Carlo results.
For now, let us slowly build our way up 
to these results, starting from a discussion of what's been done on the subject previously.

\subsection{Overview and literature review}    
    
Neutron matter is an important component in the study of neutron stars and neutron-rich nuclei~\cite{Gandolfi:2015}, so it has received a lot of attention. It has been tackled via a variety of \textit{ab initio} many-body approaches~\cite{Friedman:1981,Akmal:1998,Schwenk:2005,Gezerlis:2008,Epelbaum:2008b,Kaiser:2012}. Like the two-neutron system, many neutrons do not form bound systems; neutron matter in its simplest incarnation is a homogeneous and isotropic fluid. It can only exist under rather extreme conditions of pressure and density as found in a neutron star. Even then, the neutron matter present is not homogeneous due to nuclei in the crust. 
 
    Central to our understanding of nuclear phenomena is the interplay of empirical data (experimental and astrophysical) and theory. This is an exciting time, due to the recent measurement of a gravitational-wave signal coming from a neutron-star merger~\cite{LIGO}. The bulk of empirical information employed by nuclear theory comes from studies of nuclei. The most universal calculations for general nuclear phenomena are those utilizing nuclear energy-density functionals (EDFs)~\cite{Bender:2003}. EDFs contain parameters that are constrained using empirical data and/or \textit{ab-initio} many-body calculations. Examples of such constraints include the EOS of neutron matter~\cite{Fayans,SLy,Brown:2000,Gogny,Fattoyev:2010,Fattoyev:2012,Brown:2014,Rrapaj:2016}, the neutron pairing gap~\cite{Chamel:2008}, the neutron polaron~\cite{Forbes:2014,Roggero:2015}, investigations of neutron drops~\cite{Pudliner:1996,Pederiva:2004,Gandolfi:2011,Potter:2014}, and the static-response problem~\cite{Pastore:2015,Chamel:2014,Davesne:2015,Buraczynski:2016,Buraczynski:2017,Boulet:2018}. In addition to EDFs, neutron systems provide an excellent setting for testing state-of-the-art nuclear forces. These can be both phenomenological~\cite{Carlson:Morales:2003,Gandolfi:2009,Gezerlis:2010,Gandolfi:2012,Baldo:2012} and chiral~\cite{Hebeler:2010,Gezerlis:2013,Coraggio:2013,Hagen:2014,Gezerlis:2014,Carbone:2014,Roggero:2014,Wlazlowski:2014,Soma:2014,Tews:2016,Piarulli:2018,Lonardoni:2018}.
    
In this paper we first review our previous work on the static-response problem~\cite{Buraczynski:2016,Buraczynski:2017}. The response problem is a comparison between unperturbed and perturbed neutron matter. Other studies have been conducted approximating such calculations in nuclear systems~\cite{Iwamoto:1982,Olsson:2004,Chamel:2011,Chamel:2013,Kobyakov:2013,Pastoretheory,Chamel:2014,Davesne:2015}.  Ref.~\cite{Pastore:2015} is a review of response in nuclear matter. Static response has a long history outside
nuclear physics~\cite{Pines:1966}. Noteworthy contributions include early QMC calculations in zero-temperature-and-pressure liquid $\rm {{}^4He}$~\cite{Moroni:1992} as well as the three-dimensional electron gas~\cite{Moroni:1995}.
        In our own work, we employed two complementary approaches: QMC and nuclear energy density functionals. The term ``static response'' clarifies that the impact on the time-independent energy eigen-states are studied as opposed to system evolution under a time-dependent perturbation. We study neutron matter at zero temperature where all the properties examined are ground state.
        In the spirit of the present paper, we emphasize how an EDF parameter can be matched onto the microscopic QMC results,
        for a case where no experimental input is available.
        
        The primary investigations in this paper are on the quasiparticle energy dispersion relation at the Fermi surface. This is relevant to one of the parameters used in nuclear EDFs: the effective mass in neutron matter. We provide several new derivations
        and plots, expanding on our work published in Ref.~\cite{Ref:our_first_paper_effmass}. A thorough review on effective masses in neutron-rich matter is Ref.~\cite{Li:2018}. The basic notion behind the effective mass aims to treat the interactions of a particle amongst others by introducing the notion of a \emph{quasiparticle}. Intuitively, this quasiparticle can be thought of
        as encompassing the bare mass along with the interactions of neighboring particles into a sort of ``cloud''. Then this quasiparticle can be thought of as a free particle to leading order (since the interactions are now in its cloud).
 The effective mass impacts important quantities in nuclear physics including thermodynamic properties like the thermal index, the maximum mass of a neutron star, the static response of nucleonic matter and analyses of giant quadrupole resonances.
       Given its importance, it comes as no surprise that there have been several extractions of the effective mass using various many-body methods~\cite{Friedman:1981,Wambach:1993,Schwenk:2003,Drischler:2014,Isaule:2016,Grasso:2018,Bonnard:2018}. However, the results do not give a consistent answer, since
       different approaches make different approximations and assumptions about what the effective mass
       means (e.g., effective mass in the single-particle spectrum).       
        Our calculations provide a systematized, model-independent extraction of the effective mass in an attempt to resolve this question.
        
        Ref.~\cite{Ref:our_first_paper_effmass} was the first reference to do QMC calculation extractions of the effective mass, here we provide several new calculations and insight. To extract the effective mass, we make use of Auxiliary Field Diffusion Monte Carlo (AFDMC) to perform energy calculations for several excited states with which we can probe the quasiparticle dispersion relation. This can be used to capture the influence of the interactions by comparing to the quadratic dispersion relation for the free particle. Additionally, we wish to extrapolate our finite $N$ calculations to the infinite system (since this is representative of realistic neutron matter). To do this we apply an extrapolation prescription which attempts to minimize the effects of the periodic boundary conditions; as part of this, we provide an original derivation for the kinetic energy, which we then
        employ as a correction term. 
        
        Combining these two methodologies, we take pains to select the ideal particle number from which we can extract our results by performing a systematic review of the $N$ dependence on the energies involved. With this, we determine the general trend that the effective mass at low densities approaches unity as it acts more like the free system, while at higher densities a steady decrease is found. We try to interpret this finding qualitatively.

\section{Methods}\label{sec:methods} 
    \subsection{Hamiltonian}\label{sec:Hamiltonian}
        The many-body neutron system is modelled using a non-relativistic nuclear potential containing two and three-body forces:
        \begin{align}
        \hat{H}=-\frac{\hbar^2}{2m}\sum_{i}{\nabla_i^2}+\sum_{i<j}{v_{ij}}+\sum_{i<j<k}{v_{ijk}},
        \label{eq:I_H}
        \end{align}
      Higher-order many-nucleon forces exist as well, though the four-body forces and beyond are small. There is no unique nuclear Hamiltonian. In our work we have utilized both phenomenological potentials and effective field theories~\cite{Gandolfi:2015}.
     
     A wealth of neutron-proton scattering data has resulted in several high-quality phenomenological formulations of the nuclear potential. The Argonne family of two-body potentials takes on an operatorial structure of radial functions multiplying spin, tensor, spin-orbit, isospin, and several other operators. The Argonne v8' (AV8') potential is employed in our calculations~\cite{Wiringa:2002}. It has the form:
\begin{align}
v^{NN}&=\sum_{i<j}v_{ij} \nonumber \\
v_{ij}&=\sum_{p=1}^8v_p(r_{ij})\hat O_{ij}^p
\label{eq:Av8'}
\end{align}
where $r_{ij}$ is the inter-particle distance and
\begin{align}
\hat O_{ij}^{p=1,8}=(1,\boldsymbol\sigma_i \cdot \boldsymbol\sigma_j,S_{ij},\mathbf{L}_{ij}\cdot{\boldsymbol\sigma}_{ij})\otimes(1,\boldsymbol\tau_i \cdot \boldsymbol\tau_j)
\label{eq:Argonneop}
\end{align}
For neutron matter the isospin components coming from $\boldsymbol\tau_i \cdot\boldsymbol\tau_j$ can be trivially handled. We are left with four terms: a central potential, a spin-spin term, the tensor term $S_{ij}$ which depends on both the inter-particle separation vector and the spin of the particles, and the spin-orbit term. To be more specific:
\begin{align}
    &S_{ij}=3(\boldsymbol\sigma_i\cdot\hat{\mathbf r}_{ij})(\boldsymbol\sigma_j\cdot\hat{\mathbf r}_{ij})-\boldsymbol\sigma_i \cdot \boldsymbol\sigma_j, \nonumber\\
    &\mathbf L_{ij}=\frac{\hbar}{2i}(\mathbf r_i-\mathbf r_j)\times(\boldsymbol\nabla_i-\boldsymbol\nabla_j),\,\, \rm{and}\nonumber\\
    &\boldsymbol\sigma_{ij}=\frac{\hbar}{2}(\boldsymbol\sigma_i+\boldsymbol\sigma_j)\nonumber
\end{align}
In the spirit of Eq.~(\ref{eq:I_H}) a phenomenological three-body interaction is used in addition to AV8'. We employ Urbana-IX (UIX)~\cite{Pudliner:1997} which was fit to light nuclei and nuclear matter when it was designed.

We will also present results produced using chiral effective field theory interactions. Chiral EFT is based on the symmetries of Quantum Chromodynamics (QCD). The theory systematically expands the force via a power counting scheme. The expansion employs a separation of scales which is the ratio of the pion mass (nucleonic scale) to a hard momentum scale where the theory is expected to break down. The terms in the expansion are then given in powers of this expansion parameter. The many-nucleon forces arise naturally in chiral EFT. They include known pion exchanges as well as phenomenological short-range terms. The expansion terminology is: leading-order (LO), next-to-leading order (NLO), next-to-next-to-leading order (N$^2$LO) etc. Note that three-body forces do not appear until N$^2$LO. We consider both the chiral forces truncated to NN from~\cite{Gezerlis:2014} and those including NNN introduced in~\cite{Tews:2016}.
        
        \subsection{Quantum Monte Carlo}\label{sec:QMC}
        The task of solving Schr\"odinger's Equation for any state-of-the-art nuclear potential is a highly demanding computational problem. We approach it with stochastic methods called quantum Monte Carlo (QMC)~\cite{Pudliner:1997,Gandolfi:2012,Gezerlis:2013}. Specifically, we use AFDMC, which is an extension of a projection algorithm called Diffusion Monte Carlo (DMC). DMC projects the ground state out from a trial wave function $|\Psi_T \rangle$ by the imaginary time evolution:        
        \begin{align}
            |\Phi_0 \rangle= \lim_{\tau \to \infty} |\Phi(\tau)\rangle=\lim_{\tau \to \infty} e^{-(\widehat{H}-E_T)\tau}|\Phi(0)\rangle.
            \label{eq:prop}
            \end{align}
                        The trial wave function is the initial state of the evolution: $|\Psi_T\rangle=|\Phi(\tau=0)\rangle$. $E_T$ is simply a normalizing energy offset that prevents the evolving state from vanishing or blowing up. 
            
            The Trotter-Suzuki trick allows us to write the projection in imaginary time as many integrals over coordinate space where each integral corresponds to an evolution over a sufficiently small time step. This small time step evolution allows us to analytically approximate the matrix elements of the imaginary time propagator. Monte Carlo itself is the stochastic method by which these integrals are evaluated. 
            
            While QMC is in principle an exact method, complications arise when dealing with a fermionic wave function. Anti-symmetry forces the wave function to have zeros in coordinate space. The change of sign of the wave function is called the fermion sign problem because plain Monte Carlo integration works with functions of a single sign. One of the primary methods to deal with this is to fix the nodal surface or something equivalent in the case of complex wave functions. It has been shown that this approach still yields accurate results when the trial wave function is intelligently designed to encapsulate some of the ground-state physics. We use a trial wave function that is a product of a Slater determinant with a nodeless Jastrow factor:            
            \begin{align}
            |\Psi_T \rangle =\prod_{i<j}f(r_{ij})\,\,\mathcal{A}\bigg[\prod_i|\phi_i,s_i\rangle\bigg].
            \label{eq:trial}
        \end{align}
                We start by using Variational Monte Carlo (VMC); the expectation value of the energy is given by:
        \begin{align}
E&=\int\Bigg(\frac{|\Psi_T(\mathbf{R})|^2}{\int|\Psi_T(\mathbf{R})|^2 d\mathbf R}\Bigg)\frac{\hat{H}\Psi_T(\mathbf{R})}{\Psi_T(\mathbf{R})}d \mathbf{R} \nonumber \\
&=\int P_T(\mathbf{R})E_L(\mathbf{R})d\mathbf{R}
\label{eq:energy2}
\end{align}
In VMC $P_T$ is sampled and $E_L$ is averaged to evaluate the integral. By introducing a variational parameter, we can optimize the trial wave function. This also serves to obtain initial configurations for the imaginary time evolution.
        
        For our nuclear interactions, the specific calculations are performed with Auxiliary Field Diffusion Monte Carlo  because it is capable of handling the complicated spin dependence that arises from those forces~\cite{Schmidt:1999}. This is accomplished through the use of the Hubbard-Stratonovich transformation and reduces the number of operations involving spin from scaling exponentially to linearly in the number of particles.
        The overall complexity of AFDMC calculations scales as the cube of the number of particles being simulated. This is mainly due to wave function evaluations. This limits simulations to about $100$ particles. The final output is an estimation of the ground-state energy of the system.
    
\section{Matching EDFs to QMC: A Summary}\label{sec:response}
Given the theme of the present paper, namely the interface between different approaches to the nuclear many-body
problem, it may be appropriate to first summarize some of our recent work from Refs.~\cite{Buraczynski:2016} and~\cite{Buraczynski:2017}. We consider the static-response due to a one-body potential:
        \begin{align}
        &\hat{H}=-\frac{\hbar^2}{2m}\sum_{i}{\nabla_i^2}+\sum_{i<j}{v_{ij}}+\sum_{i<j<k}{v_{ijk}}+v_{\rm ext},\nonumber\\
            &v_{\rm ext}=\sum_{i}{v(\mathbf{r}_i)}~\text{and}~v(\mathbf{r}_i)=2v_{q}\cos(\mathbf{q}\cdot \mathbf{r}_i).
        \end{align}
        The significance of this choice is twofold. A periodic modulation is useful for describing inhomogeneous neutron matter. One such system with an albeit more complicated modulation is the neutron fluid inside of a neutron-star crust that is perturbed by a lattice of neutron-rich nuclei. Secondly, calculations with a monochromatic potential can be used to extract the linear static response function of a system with relative ease~\cite{Moroni:1992,Buraczynski:2017}.
        
        In addition to performing QMC calculations for this response, we contrasted these results with energy density functional calculations.
        In the latter, many-body properties are expressed in terms of the one-body density. In nuclear physics, EDFs are often derived via the use of effective interactions. Fundamental physics is exchanged for computational ease and the ability to study larger nuclear systems. This is achieved, at least in part, by parameterizing and fitting components of the interaction to nuclear data.
        
        \begin{figure}
            \resizebox{0.485\textwidth}{!}{%
              \includegraphics{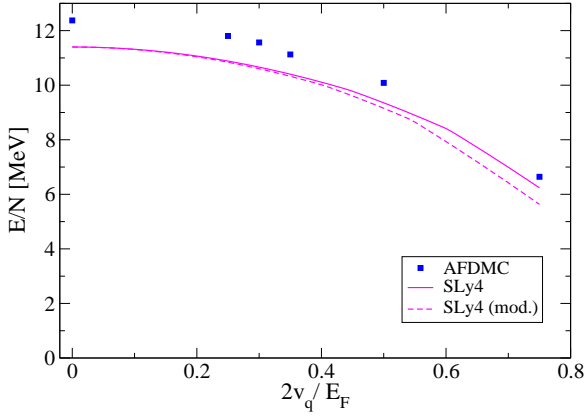}
            }
            \caption{Energy per particle for 66 neutrons, displaying the response to increasing strengths of the one-body potential at a density of $0.10\, \rm{fm}^{-3}$. Two periods of the potential span the box containing the particles. The squares are AFDMC calculations from two-body AV8' and three-body Urbana IX forces. The curves are EDF calculations. The solid line is SLy4 whereas the dashed line is SLy4 modified by fitting the isovector coefficient to match the QMC response. Results taken from Ref.~\cite{Buraczynski:2017}.
            \label{fig:ener_v_stren}}
        \end{figure}

        We studied effective interactions of the Skyrme type. This is a zero-range effective interaction which yields algebraic functional dependence that is easy to work with. The many-body wave function is described by a Slater-determinant of single-particle orbitals. The relevant one-body densities for neutron matter calculations are the nucleon number and kinetic energy densities:
        \begin{align}
            &n(\mathbf{r})=\sum_i[\psi_i(\mathbf{r})]^2 \nonumber \\
            &\tau(\mathbf{r})=\sum_i[\nabla\psi_i(\mathbf{r})]^2
            \label{eq:rho}
        \end{align}
        calculated from the single-particle $\psi_i$ orbitals. The energy can be shown to be given by
        \begin{align}
            &E=\int \mathcal{H}(\mathbf{r})d^3r
            \label{eq:DFT1}
        \end{align}
                where $\mathcal{H}$ is called the energy density functional~\cite{Bender:2003}.
                \begin{align}
            \mathcal{H} = \frac{\hbar^2}{2m}\tau+2v_q\cos(\mathbf{q}\cdot \mathbf{r}) n+{\cal E}_{Sk}
            \label{eq:DFT2}
        \end{align}
                The second term in Eq.~(\ref{eq:DFT2}) comes from the external perturbation. The last term contains the Skyrme interactions which take the form:
                \begin{align}\label{eq:EDF3}
            {\cal E}_{Sk} = \sum_{T=0,1} \big [
                ( C^{n,a}_T + &C^{n,b}_T n^{\sigma}_0  ) n^2_T +\\ 
                 &C^{\Delta n}_T (\nabla n_T)^2 + C^{\tau}_T n_T \tau_T
             \big ] \nonumber
        \end{align}
                in the isospin representation. All the densities are just the corresponding neutron density in the case of pure neutron matter. We have performed calculations for three different Skyrme parameterizations: SLy4, SLy7, and SkM*. Of these we mainly focused on SLy4. The $C^{\Delta n}_1$ parameter is called the isovector gradient term. We shall show how we can tune this coefficient based on our AFDMC calculations.
        
\begin{figure}
            \resizebox{0.485\textwidth}{!}{%
              \includegraphics{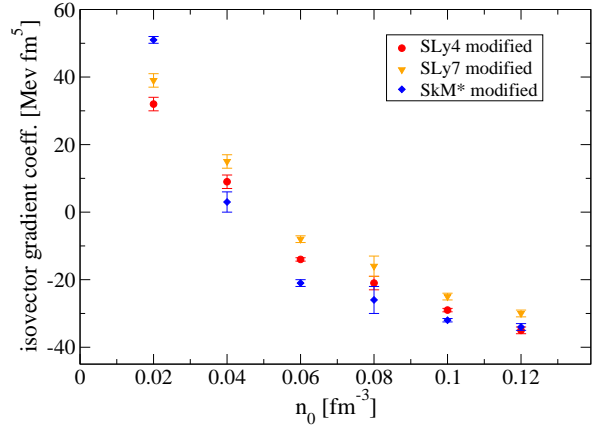}
            }
            \caption{Modified isovector coefficients after matching the SLy4, SLy7, and SkM* effective interactions onto the QMC response. Results are for 66 particles and two periods of the potential in the box. Unmodified values are $C^{\Delta n}_1=-16$, $-6$, and $-17$ MeV fm$^{5}$ for SLy4, SLy7 and SkM* respectively~\cite{SLy}. Results taken from Ref.~\cite{Buraczynski:2017}.\label{fig:3funcs}}
        \end{figure}

        We attack Eq.~(\ref{eq:DFT1}) using a variational approach. Rather than employing a self-consistent Hartree-Fock approach, we limit the space of our one-particle orbitals and minimize the energy with respect to a variational parameter. 
        The calculations we performed were at the QMC ``magic number'' of 66. Despite not being so large, this closed shell of the free Fermi gas is a sweet spot for probing thermodynamic-limit physics. Note that we are dealing with 66 particles in periodic
        boundary conditions. The latter serve to capture the physics of an extended system. Our results on the static response presented in this paper are all for two periods of the cosine potential spanning the length of this box.  Neutron number densities in the range of $0.02~\rm{fm}^{-3}$ to $0.12~\rm{fm}^{-3}$ were studied as seen in Fig.~\ref{fig:3funcs}. These are motivated by the typical densities found inside of a neutron star's crust and outer core.

        The input wave functions to both our QMC and Skyrme calculations are given by antisymmetrized Mathieu functions. As solutions corresponding to a non-interacting gas with a one-body cosine external potential, this Slater determinant serves as our best guess at capturing the nodal surface requirement imposed by the fermion sign problem (QMC requirement). In the EDF approach they serve as our ``basis'' for the wave function space in which we minimize the energy. The space itself is defined by varying $v_q$ which yields a continuous spectrum of Mathieu functions. Note that these reduce to sines and cosines at $v_q=0$. The ordering of the lowest energy single-particle orbitals changes with $v_q$. This was accounted for by considering the various orderings separately, varying $v_q$ while maintaining a particular ordering, and taking the smallest energy found out of all the sets of orderings.
        
        AFDMC calculations were performed for the periodic potential with strengths, $2v_q$ of:  $0$, $0.25$, $0.3$, $0.35$, $0.5$, and $0.75$ $E_F$, where the Fermi energy is $E_F = \hbar^2 k_F^2 /(2m)$. The strengths were chosen to be large enough that the VMC calculations are statistically different from homogeneous $v_q=0$. We found that the energy per particle decreases as the potential strength increases. Fig.~\ref{fig:ener_v_stren} shows results from~\cite{Buraczynski:2017}, displaying a concave down curvature in this relationship. The calculations are for a density of $0.10\,\rm{fm}^{-3}$. The drop in energy can be interpreted as the gathering of neutrons in the wells of the cosine potential. The SLy4 results (solid line) show less curvature than the AFDMC calculations (squares). In as far as the external potential does not impact the bulk properties of our system, which is true for small perturbations, its effect is of introducing fluctuations in the one-particle density. The component of the EDF that captures the energy contribution of such density fluctuations is the isovector gradient term. Thus, AFDMC response results can be used to update the isovector gradient term in the Skyrme EDF by fitting to the energy curvature induced by the external perturbation. Any attempt to match the Skyrme energy on top of the AFDMC energy would involve the bulk parameters as well. The dashed line displays the Skyrme curvature post-fitting of the isovector gradient term. The curvature is as close to AFDMC as the single-free-parameter fit allowed. The modified isovector coefficients themselves are reported in Fig.~\ref{fig:3funcs} whose results are from~\cite{Buraczynski:2017}. The SLy4 modified isovector gradient term at $0.10\,
        \rm{fm}^{-3}$ is smaller than the unmodified value as expected from the required negative curvature in Fig.~\ref{fig:ener_v_stren}. Results are shown for SLy4, SLy7 and SkM* at various densities. There is a clear density dependence reflecting larger attractive adjustments towards higher density (Skyrme does not curve enough) and repulsive adjustments at low density (Skyrme curves more than AFDMC).

\section{Effective Mass Calculations}\label{sec:effmass}
    \subsection{Setup}\label{sec:AFDMCeffmass}
\begin{figure}
            \resizebox{.485\textwidth}{!}{%
              \includegraphics{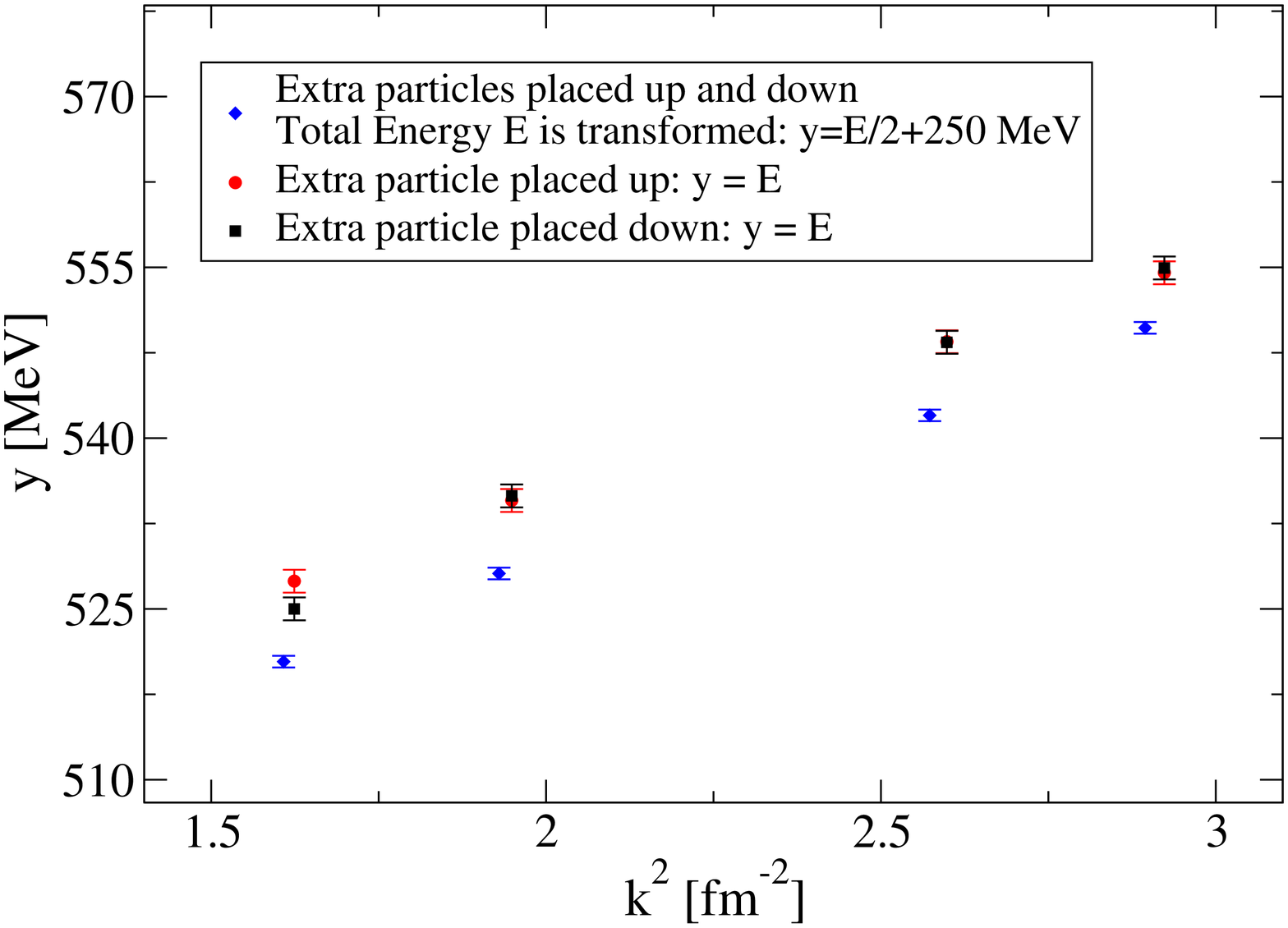}
            }
            \caption{The total interacting energies for 67 and 68 particles using the AV8'+UIX interactions at a particle density of $0.05$ fm$^{-3}$. In all three cases shown, the first 66 particles are split evenly into spin up and spin down. The red circles and black squares show the energies where the 67th particle is in the up, and down state, respectively. Finally, the blue diamonds show the 68-particle case where one additional particle is placed in each spin state. This energy is scaled, to account for the addition of energy from two particles (instead of one), as well as translated, for ease-of-viewing. All slopes (from which the effective mass could be extracted) remain the same.}
            \label{fig:spurious}
        \end{figure}

		To determine the effective mass of a particle via AFDMC calculations we must probe the energy dispersion relation, so that we may compare it to that of a free particle. Thus, we consider an excited particle. To probe the dispersion relation for a fermionic system in its ground state, we need to perturb about the Fermi surface. Previous work~\cite{Forbes:2014} has considered the  neutron polaron, in which all but one particle are of a given spin projection, but this is not a realistic system. Since neutron matter is generally not polarized~\cite{Gezerlis:2012}, we populate our system of $N$ particles evenly between the two spin states. Now, in probing the dispersion relation of a single particle, we need to decide where to place it. To investigate the effects of this, we consider placing the additional particle in spin up, and spin down. Additionally, to catch any spurious effects due to this single-particle polarization, we consider the energy of $N+2$ particles with one additional particle placed in each spin projection. Fig.~\ref{fig:spurious} shows that the slope of the dispersion relation remains unchanged. Since only the slope is required as per Sec.~\ref{sssec:emr_extraction}, we can extract the effective mass equally well by placing the additional particle in either spin projection. Arbitrarily, we choose to place it in the up state.
		
		We start with a system containing $N$ particles in the ground state, with $N/2$ in the up state, and $N/2$ in the down state. As discussed above, we consider a second system containing $N+1$ particles, with the additional up-spin particle placed at some excited momentum $k$. To probe the dispersion relation of this quasiparticle we need to remove the energy associated with the ground state, leaving us with an energy associated solely with the excited particle. In order to do that, we first
		go over some details on the free Fermi gas.
		        
     	\begin{table}[ht]
    	    \centering
    	    \caption{
    	        The degeneracies of (the integer part of) the momentum vectors for fermions of a single spin. Note that $n^2=7$ does not appear.
            }
            \label{tab:nstates}
    		\begin{tabular}{llc}
    			\hline\noalign{\smallskip}
    			Index & $n^2$ & Degeneracy \\
    			\noalign{\smallskip}\hline\noalign{\smallskip}
    			0 & 0 & 1 \\
    			1 & 1 & 6 \\
    			2 & 2 & 12 \\
    			3 & 3 & 8 \\
    			4 & 4 & 6 \\
    			5 & 5 & 24 \\
    			6 & 6 & 24 \\
    			7 & 8 & 12 \\
    			\noalign{\smallskip}\hline
    		\end{tabular}
    	\end{table}

    \subsubsection{Free Fermi gas}\label{sec:FFG}
        Consider a free Fermi gas at density $n=N/V$. The particles are in a box of length $L_N=V^{1/3}=\left(N/n\right)^{1/3}$ with periodic boundary conditions. The allowed wave-vectors are $\mathbf{k}=\left(2\pi/L_N\right)(n_x,n_y,n_z)=\left(2\pi/L_N\right){\mathbf n}$ where the $n$'s are integers. Two particles are allowed per wave-vector, due to spin degeneracy. Degenerate energies besides those from spin exist. For example, $|(1, 0, 0)|^2 = |(0, -1, 0)|^2$. Then in the ground state, the states are filled up according to these energy levels. Table~\ref{tab:nstates} lists the lowest energy levels and their single spin degeneracy. The notation $n^2$ refers to the square of the integer part of the wave-vector. We emphasize that there is no $n^2=7$ value since there is no set of three squares that sum to 7; there are infinitely many such missing values.

        The energy of an individual particle with wave vector $\mathbf{k}$ is $(\hbar^2/2m)\mathbf{k}^2$. Then the ground-state energy for a system of particles is:
        \begin{equation}
            T_N=\frac{\hbar^2}{2m}\bigg(\frac{2\pi}{L_N}\bigg)^2f(N)
            \label{eq:Etot}
        \end{equation}
        where $f(N)=\sum_{{\mathbf n}} (n_{x}^2+n_{y}^2+n_{z}^2)$ is summed over the lowest available energies. Additionally, we use the notation $T$ for the energy of the free Fermi gas, because we are reserving $E$ for the interacting problem. Keeping the density constant as the particle number grows yields the thermodynamic-limit expression:
        \begin{align}
            \lim_{N\to\infty}\frac{T_N}{N}=\frac{3}{5}E_F
            \label{eq:TLdef}
        \end{align} 
        where $E_F$, called the Fermi energy, is the highest occupied energy level. $k_F$ is the magnitude of the corresponding Fermi wave-vector and satisfies $k_F^3=3\pi^2n$ where n is the particle number density.
        We examine the energy difference associated with adding a particle to the system at constant density. We use the label $T^{(\mathbf{k})}_{N+1}$ as the total energy of $N+1$ particles with the extra particle placed at wave-vector $\mathbf{k}$. Naively one may expect the energy of the extra particle to simply be $T_{N+1}^{(\mathbf{k})}-T_{N}$. However, this does not take into account the energy difference in the first $N$ particles induced by the change in box size. This difference does not go to zero in the thermodynamic limit:
        \begin{align}
            &\frac{\hbar^2}{2m}\bigg(\frac{2\pi}{L_{N+1}}\bigg)^2f(N)-\frac{\hbar^2}{2m}\bigg(\frac{2\pi}{L_{N}}\bigg)^2f(N) \nonumber \\
            &=\frac{\hbar^2}{2m}f(N)(2\pi)^2n^{2/3}\bigg(\frac{1}{(N+1)^{2/3}}-\frac{1}{N^{2/3}}\bigg) \nonumber \\
            &=\frac{\hbar^2}{2m}f(N)(2\pi)^2n^{2/3}\sum_{i=1}^{\infty}\frac{1}{i!}\frac{d^{i}}{dx^i}\bigg(\frac{1}{x^{2/3}}\bigg)\bigg|_{x=N} \nonumber \\
            &= \frac{\hbar^2}{2m}f(N)(2\pi)^2n^{2/3}\sum_{i=1}^{\infty}(-1)^i\frac{\prod_{j=1}^i(3j-1)}{3^ii!N^i}\frac{1}{N^{2/3}} \nonumber \\
            &=\frac{T_N}{N}\sum_{i=1}^{\infty}(-1)^i\frac{\prod_{j=1}^i(3j-1)}{3^ii!}\frac{1}{N^{i-1}} \nonumber \\
        \end{align}
                 where we used Eq.~(\ref{eq:Etot}) to get from the fourth line to the fifth. Taking the thermodynamic limit and applying Eq.~(\ref{eq:TLdef}) to this result yields an overall energy difference of $-(2/5)E_F$.
        
        Analogous arguments hold for the case of the interacting gas.
		The naive approach would be simply:
		\begin{equation}
			\Delta E^{(k)} \equiv E_{N+1}^{(k)}(N+1) - E_{N}(N),
		\end{equation}
		where $E_{N+1}^{(k)}(N+1)$ represents the energy of all $N+1$ particles in a box containing $N+1$ particles, with the last being in an excited state, $k$. Similarly, $E_{N}(N)$ is the energy of the $N$ ground-state particles in a box containing $N$ particles. However, we see that this energy does not properly extract the energy we want:
		\begin{align}
			\Delta E^{(k)} &= E_{N+1}^{(k)}(1) + E_{N+1}(N) - E_{N}(N) \nonumber\\
			&= E_{N+1}^{(k)}(1) -\frac{2}{5}\xi E_F,
			\label{eq:Ediff}
		\end{align}
		where $\xi$ is a factor relating the interacting energy per particle to the free energy per particle. $E_{N+1}^{(k)}(1)$ and $E_{N+1}(N)$ separates the excited particle's energy from the remaining particles in the $N+1$ system. The above definition has an offset which is undesirable, so we redefine $\Delta E^{(k)}$ to be,
		\begin{equation}
			\Delta E^{(k)} \equiv E_{N+1}^{(k)} - E_{N} + \frac{2}{5}\xi E_F,
			\label{eq:DeltaE}
		\end{equation}
		where we dropped the parentheses since in this context, they are redundant. We can consider a twin equation for the non-interacting system,
		\begin{equation}
			\Delta T^{(k)} \equiv T_{N+1}^{(k)} - T_{N} + \frac{2}{5}E_F.
			\label{eq:DeltaT}
		\end{equation}
        
        \subsubsection{Effective Mass Extraction}\label{sssec:emr_extraction}
Armed with these definitions, we now turn to the concept of the effective mass. 
We start from the dispersion relation of a non-interacting, excited particle:
    		\begin{equation}
    			\Delta T^{(k)} = \frac{\hbar^2}{2m}k^2.
    		\end{equation}
    		In the literature, what is usually done at this point is an expansion around the Fermi surface:
\begin{equation}
\frac{\hbar^2}{2m}k^2 - E_F \approx \frac{\hbar^2 k_F}{m} (k - k_F)
\label{eq:freedisp}
\end{equation}    		
This is legitimate as long as $k$ is very close to the Fermi surface, i.e., $k - k_F \ll k_F$. However,
note that this is a further approximation, which is not necessary to introduce the intuitive concept of quasiparticles;
the latter intuitive picture arises in the interacting case, to which we now turn.
    	
    		 \begin{figure}
            \resizebox{0.485\textwidth}{!}{%
              \includegraphics{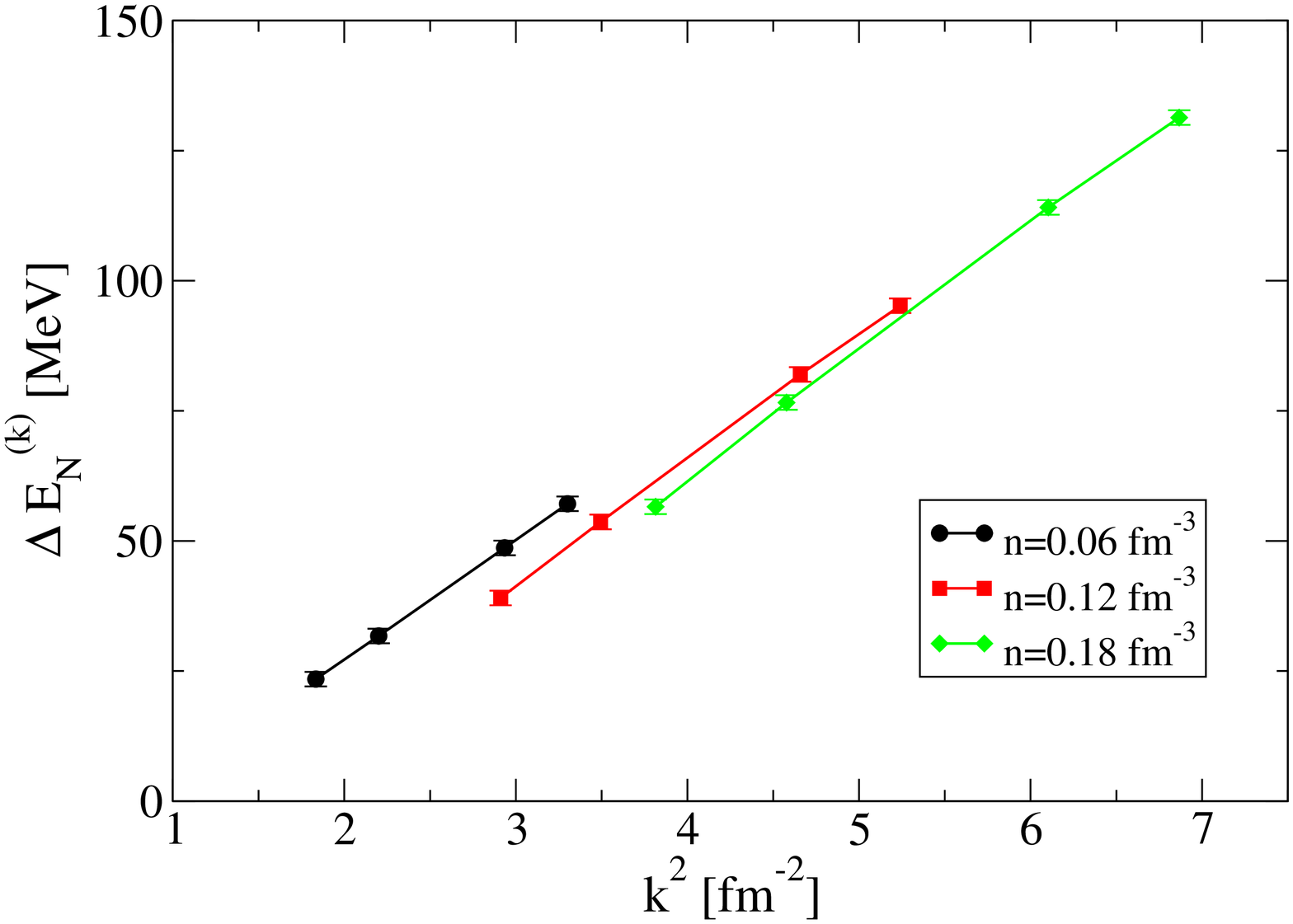}
            }
            \caption{
                Quasiparticle energies for neutron matter as a function of the excited particle's squared momentum $k^2$. These energies are calculated using the AV8'+UIX interactions, carried out with AFDMC at densities of 0.06, 0.12, and 0.18 fm$^{-3}$. The momentum for finite (non-superfluid) Fermi systems comes in discrete spacings meaning that although the excited states are not infinitesimally close, these curves are still the energy dispersion relations near the Fermi surface.
                \label{fig:fig_deltaE_vs_k2}
            }
        \end{figure}
		
Motivated by Eq.~(\ref{eq:freedisp}), it is customary to introduce an effective mass $m^*$ for the interacting
problem, where the right-hand side will involve $m^*$ instead of $m$. However, as we just pointed out, this is only
legitimate if $k - k_F \ll k_F$ holds. In our QMC simulations we do not have $k$ at our disposal: this is determined
by the density and particle number. Thus, we keep things general, i.e., we introduce the effective mass still at the level
of the quadratic dispersion, as in the left-hand side of Eq.~(\ref{eq:freedisp}). To be explicit, for the interacting
problem we introduce a parameter $m^*$ that is intended to capture the effects of the interactions on the quasiparticle dispersion:
    		\begin{equation}
    			\Delta E^{(k)} = \frac{\hbar^2}{2m^*}k^2
    		\end{equation}
Since we still wish to remain as close to the Fermi surface as possible, we have chosen 4 points as a happy medium between 
having several input data points and not straying too far away from the surface. 
In what follows, we will fit our QMC results to such a quadratic form in order to extract the effective mass.
    		
    		Before we can extract the coefficient $m^*$, we need to produce microscopic results of $\Delta E^{(k)}$ 
    		as a function of the squared excitation momentum, $k^2$. 
    		Ref.~\cite{Ref:our_first_paper_effmass} reported on results at a single density; here we expand on this to better understand the density dependence of these quantities. In Fig.~\ref{fig:fig_deltaE_vs_k2} the dependence on the squared excited momentum is linear at a variety of densities, justifying that a linear slope is sufficient to determine the effective mass. This in turn allows us to encapsulate the many-body physics of several interacting particles, into a single-particle quantity, the effective mass. Before we can determine these ratios as a function of density, for example, we need to determine which particle numbers we trust for the extractions.

	\subsection{Energy Extrapolation}
	\label{sec:enerextrap}
	    Despite being limited to finite $N$-calculations by computational complexity, we wish to determine thermodynamic-limit (TL) interacting quantities. One prescription that we can employ is to make use of the non-interacting system to better approximate the kinetic contribution. This becomes evidently useful for neutron matter since the finite-size effects (FSE) have dominant contributions from the kinetic energy (due to a small effective range of $\sim$2.7 fm). This prescription can be carried out as follows:
		\begin{align}
			\bar E_{\infty} &= \bar V_{\infty} + \bar T_{\infty} \\
			 &\approx \bar  V_{N} + \bar T_{\infty} \\
			\bar E_{TL} &\equiv \bar E_{N} - \bar T_{N} + \bar T_{\infty},
			\label{eq:prescription}
		\end{align}
		where symbols with a bar over them represent energies per particle, and we let $V, T,$ and $E$ denote the potential energy, free particle energy (or kinetic energy), and interacting energy, respectively. We use energies per particle since we will be applying this to different particle numbers, and this makes the last term a finite quantity. Although this significantly lowers FSE contributions, the left-hand side still depends on $N$ and too low a particle number will produce poor results. We use $N=66$ as a starting point in our calculations. Our first application of this will be in the computation of $\xi$. The definition of $\xi$ is $\bar E_{\infty}=(3/5)\xi E_F$. Thus
		\begin{align}
		    \xi &= \bar E_{\infty} / \bar T_{\infty} \nonumber\\
		  &\approx \bar E_{TL}/\bar T_{\infty}.
		\end{align}

		\begin{figure}[ht]
		    \centering
            \resizebox{0.4\textwidth}{!}{%
              \includegraphics{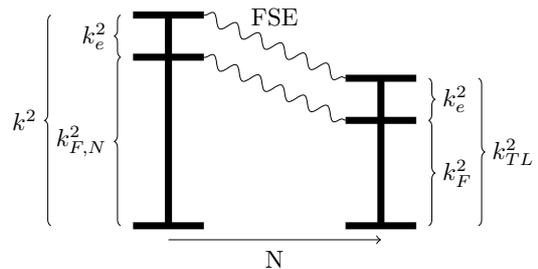}
            }
            \caption{
                The energy of an excited particle in a finite system is proportional to $k^2$ (left). This can be split into two contributions. The $N$ particle ``Fermi surface" contributes $k_{F,N}^2$ (bottom left). The remaining energy is the excitation amount $k^2_e$ (top left). In taking the TL the Fermi surface settles at $k_F$ (bottom right) as FSE go to zero. Tracking the FSE is accomplished by keeping excitation energy constant (top right). The total energy of the excited particle in the TL is proportional to $k_{TL}^2$ (right). The mathematical relationship defining $k_{TL}$ is given in Eq.~(\ref{eq:k2_TL}).
            \label{fig:k2_TL}}
        \end{figure}

		We apply the extrapolation prescription to reduce the FSE in the effective mass calculations. Consider the following:
		
		\begin{align} 
			\Delta E^{(k)}_{TL} &= [E_{N+1}^{(k)}- T_{N+1}^{(k)}+(N+1)\bar T_{\infty}^{(k)}]\nonumber\\
			&-[E_N^{(k)}-T_N^{(k)}+(N)\bar T_{\infty}]+\frac{2}{5}\xi E_F.
			\label{eq:extrapolation}
		\end{align}
		Eq.~(\ref{eq:extrapolation}) applies Eq.~(\ref{eq:prescription}) to $N$ and $N+1$ particle results. The corrected energies are at the same finite particle numbers but nevertheless treated as TL quantities since the FSE in the energy per particle have been handled. We pay close attention to the kinetic energy correction for $N+1$ particles. Without an excitation, the TL correction in energy per particle after subtracting $T_{N+1}$ is simply $(N+1)(3/5)E_F$. However, we must carefully consider the corresponding correction when a particle is placed in an excited state $k$. There is a FSE associated with the Fermi surface wave-vector $k_F$ due to the limited amount of allowed wave-vectors. The $N$ least energetic particles are all placed in the lowest available energy levels up to what we label as $k_{F,N}$. The energy level is proportional to the square of the wave-vector. We handle the movement of the Fermi surface by keeping the energy difference from the surface to the excited state constant: i.e. the excited state in the TL is not at $k$. Rather 
		
		\begin{align}
		k^2-k_{F,N}^2=k_{TL}^2-k_F^2,\,\,\,{\rm so}\nonumber\\
		k^2_{TL} = k^2 - k_{F,N}^2+k_F^2
		\label{eq:k2_TL}
		\end{align}
		where $k_{TL}$ is where the excited particle is placed in the TL. This is illustrated in Figure~\ref{fig:k2_TL}.
	
		Returning to Eq.~(\ref{eq:extrapolation}) we notice that $\bar T_{\infty}^{(k)}$ should actually be $T_{\infty}^{(k_{TL})}$. Rather than $(N+1)(3/5)E_F$, the correction is
		\begin{align}
		(N+1)\bar T_{\infty}^{(k_{TL})}=(N+1)\frac{3}{5}E_F+\frac{\hbar^2}{2m}k_{TL}^2-E_F
		\end{align}
	 where the last two terms reflect the excitation energy of the particle above the Fermi surface. Plugging this back into Eq.~(\ref{eq:extrapolation}) yields
	 \begin{align}
	 	\Delta E^{(k)}_{TL} &= [E_{N+1}^{(k)}- T_{N+1}^{(k)}+(N+1)\frac{3}{5}E_F+\frac{\hbar^2}{2m}k_{TL}^2-E_F]\nonumber\\
			&-[E_N^{(k)}-T_N^{(k)}+(N)\frac{3}{5}E_F]+\frac{2}{5}\xi E_F\nonumber\\
			&=\Delta E^{(k)} - \Delta T^{(k)} + \frac{\hbar^2}{2m}k_{TL}^2
			\label{eq:extrafinal}
	 \end{align}
	where we also plugged $\bar T_{\infty}=(3/5)E_F$ into Eq.~(\ref{eq:extrapolation}) for the first step of this derivation. Eq.~(\ref{eq:DeltaE}) and Eq.~(\ref{eq:DeltaT}) were used to obtain the final expression. We extract the effective mass from the extrapolated excited particle energy via:
		\begin{equation}\label{eq:delta_E_TL}
			\Delta E^{(k_{TL})}_{TL} = \frac{\hbar^2}{2m^*}k^2_{TL}.
		\end{equation}
	    It is quite rewarding to see this compact notation connecting the bare mass, which appears in the extrapolation process in Eq.~(\ref{eq:extrafinal}),
	    to the effective mass.
	    
\begin{figure}
                    \resizebox{0.485\textwidth}{!}{%
                      \includegraphics{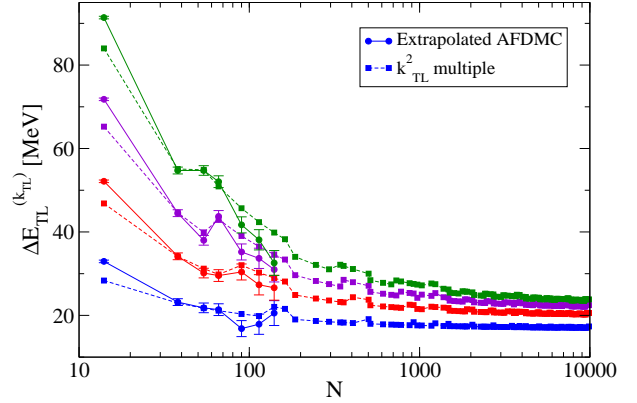}
                    }
                    \caption{
                        Guidance on the $N$-dependence of $\Delta E_{TL}^{(k_{TL})}$. These extrapolated quasiparticle energies are calculated using AFDMC and the AV8'+UIX potential at a density of 0.05 fm$^{-3}$. The first four excited state energies are shown in solid lines (from bottom to top). The dotted lines are generated by finding a coefficient which multiplies $k^2_{TL}$ to best match the quasiparticle energies over all $N$ for which AFDMC calculations were possible. Our extrapolated AFDMC values are limited to $\sim$100 particles; on the other hand, since the multiples are calculated according to the free particle momentum, they can be calculated for very large $N$. We find that for intermediate values of $N$, a multiplicative constant is sufficient to not only capture all the $N$ dependence of $k^2_{TL}$, but also to match the $\Delta E_{TL}^{(k_{TL})}$. Then, in those cases, a coefficient (the effective mass) is sufficient to convert the quadratic momentum dependence to the desired energy.
                    \label{fig:minimize}}
                \end{figure}

	    Our extraction of the effective mass was performed via a linear fit between the quasiparticle energy and the wave-vector squared. The effective mass only depends on the slope of this fit. Looking at $\Delta E^{(k)}$, we see that the only quantity that changes as $k$ changes is $E^{(k)}_{N+1}$. Similarly for $\Delta T^{(k)}$ only $T^{(k)}_{N+1}$ changes. Finally, $k^2_{TL}$ only gets an influence from $k^2$. For the latter two quantities, the $k$ dependence goes as $(\hbar^2/(2m)) k^2$ and cancels out. It is evident that the linear slope in an un-extrapolated fit of $\Delta E^{(k)}$ to $k^2$ will yield the same slope/effective mass as the extrapolated fit between $\Delta E_{TL}^{(k_{TL})}$ and $k_{TL}^2$. Nevertheless, the full quasiparticle energy quantities provide insight into how the TL is reached as seen in Fig.~\ref{fig:minimize}. This is further discussed in the next section.
	   
	   \subsection{Results}
            \subsubsection{Optimal Particle Number}
        	   Often the $N=66$ particle number is ideal for QMC calculations. It is a closed-shell (implying no ambiguity in the wave vectors used), it exhibits a minimum in FSE for bulk energies, and is a small enough particle number that we can carry it out, but not so high that we do not trust the numerical accuracy due to insufficient simulation time. However, we seek to determine the best $N$ to study since FSE may be unpredictable and minima in FSE for one observable do not necessarily imply minima in another (e.g., bulk quantities like total energy vs single-particle quantities). To carry out this analysis, we consider the extrapolated $\Delta E_{TL}^{(k_{TL})}$ as a function of particle number at a fixed density (chosen to be 0.05 fm$^{-3}$). Although ideally we would study closed-shells, their fixed spacing makes it difficult to capture the details of an $N$-dependence, so we include some open-shells as well. It is not immediately obvious how the $k$ states should be chosen for open-shells. One could place the excited particle starting at the partially filled energy level and then at several larger levels or one could start at the next lowest energy level. We carried out the latter to be consistent with closed-shell calculations. The data points that we collected for our chosen shells are shown as solid lines in Fig.~\ref{fig:minimize}. To capture the overall $N$-dependence of these quasiparticle energies, we compare these to a second quantity, namely a scalar multiple of $k^2_{TL}$ for each excitation that best fits the corresponding $\Delta E_{TL}^{(k_{TL})}$. This generates a complementary set of curves that can be compared. Interestingly, at low $N$, the multiple under-predicts the energy, while at higher $N$, it over-predicts the energy. A naive expectation that the larger particle numbers should outperform smaller numbers is therefore shown to not hold true in general. We find that the \emph{intermediate} particle numbers best capture the $N$-dependence. Given that the choices of $N=38, 54, 66$ equally capture the $N$-dependence, we chose the largest of these closed-shells to be optimal. 	    
	           	       			
    			\begin{figure}
                    \resizebox{0.485\textwidth}{!}{%
                      \includegraphics{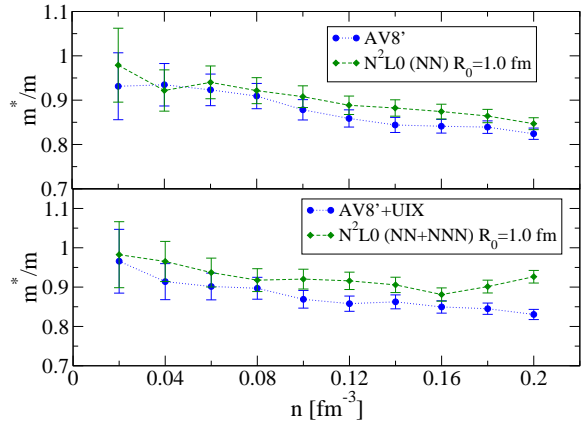}
                    }
                    \caption{
                        The effective mass ratio $m^*/m$ for neutron matter as a function of number density $n$. Results are shown for both phenomenological and chiral interactions. Two-body interaction results are shown in the top panel. The bottom panel contains calculations with three-body interactions also included. The effective mass ratio approaches unity at low densities which corresponds to large inter-particle distances causing the interacting system to act like the non-interacting system. At higher densities the effective mass decreases steadily for all potentials examined.
                    \label{fig:emr_vs_density}}
                \end{figure}

        	   Ref.~\cite{Ref:our_first_paper_effmass} provided results for $N$ up to $\sim$100. Having determined the 
        	   multiplicative coefficients, we can extend the multiples curve to much higher $N$. This now makes it clear that as we approach larger $N$, the points used in our fitting prescription come closer together. As a result, small errors in the QMC energy calculations can have significant impact on the effective mass, again reinforcing that simply increasing $N$ does not necessarily improve accuracy.
        	   

            \subsubsection{Effective Mass vs Density}
                Now that we have carried out a systematic analysis for determining the particle number which best matches the TL, and accounted for FSE, we are able to investigate the effective mass dependence on density.
      
                The results shown in Fig.~\ref{fig:emr_vs_density} employ the described fitting procedure and standard error propagation from the QMC errors. Ref.~\cite{Ref:our_first_paper_effmass} provided results for the two- and three- body Argonne and chiral potentials; we have carried out new QMC calculations at all densities, employing as input only neutron-neutron interactions (i.e., removing three-body interactions).                 At very low densities (large interparticle spacings), the interactions die off and the effective mass ratio goes
                toward 1. This is expected: as the effect of the interactions becomes vanishingly important, the dispersion relation approaches the free particle dispersion relation. 
                Since the effective mass ratio is strictly less than one for all densities, regardless of potential, neutron matter tends to gain a larger increase in energy for the same excitation than the non-interacting system. 
In order to further interpret the significance of our findings, we recall that the effective mass ratio appears in Landau Fermi-liquid theory (LFLT)~\cite{PiersColeman} in the equation:
                \begin{align}
                \frac{m*}{m}=1+F_1^s
                \label{eq:Fermiliquid}
                \end{align}
                where $F_1^s$ is the dipole component of the interactions. Note that LFLT involves an expansion
                where the coefficients are the Landau parameters. Here we see the leading spin-symmetric one, which 
                corresponds to the non-magnetic part of the interaction. The results in Fig.~\ref{fig:emr_vs_density}
                can therefore straightforwardly be used to extract $F_1^s$ in neutron matter.
                More generally, since the specific heat is proportional to the effective mass
                and the temperature~\cite{Pines:1966}, one can view our extractions as also having determined the proportionality constant between the specific heat and the temperature.
                         
                \begin{figure}
                    \centering
                    \subfigure{
                      \includegraphics[width=\linewidth]{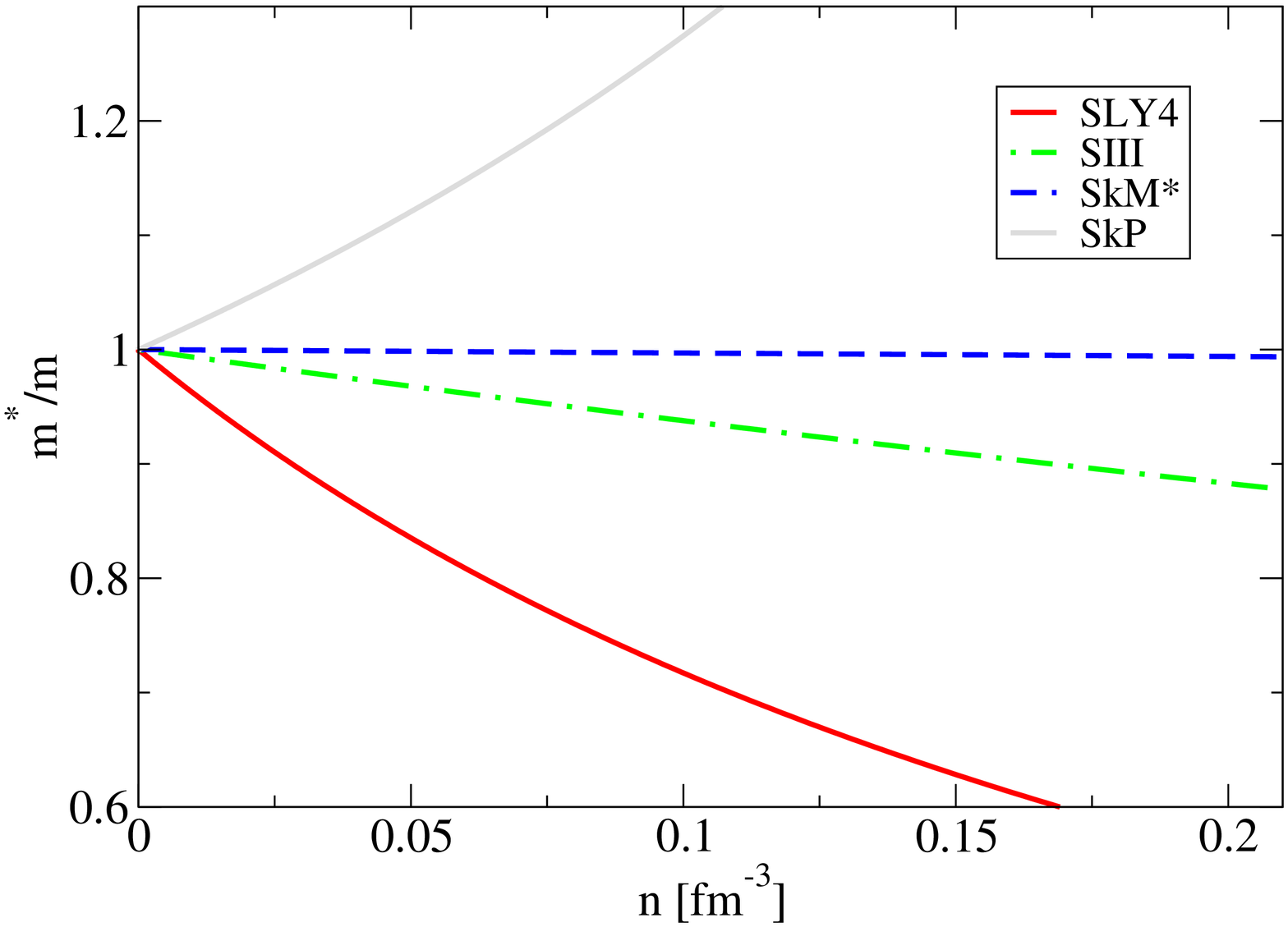}
                    }
                    \\[-5mm]
                    \subfigure{
                       \includegraphics[width=\linewidth]{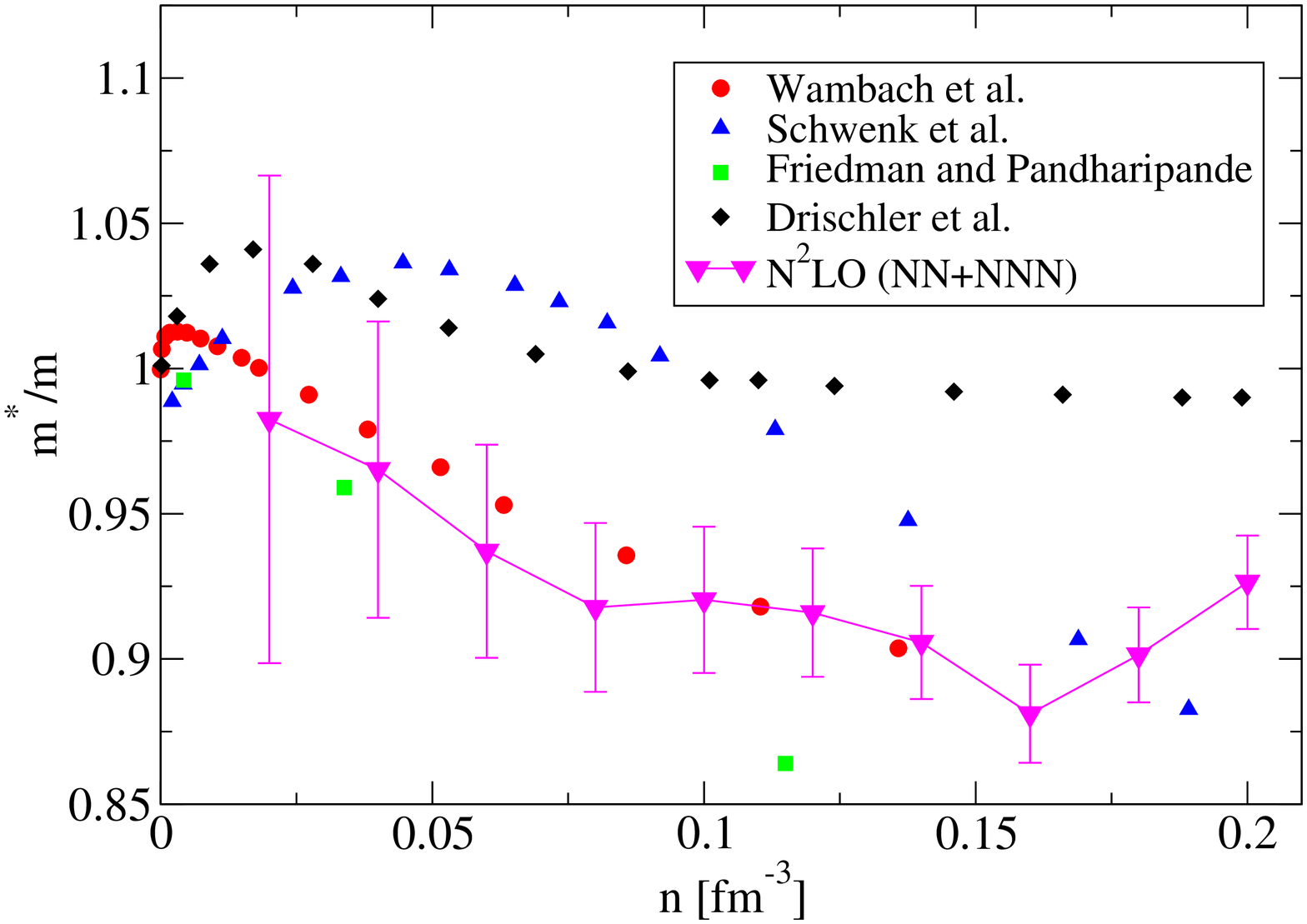}
                    }
                    \caption{Effective mass as a function of density for various approaches. The top panel shows Skyrme energy density functional results, in order from top to bottom: SLy4, SIII, SkM$^*$, and SkP. The bottom panel shows results from several \textit{ab initio} many-body calculations:  triangles \cite{Schwenk:2003}, circles \cite{Wambach:1993}, squares \cite{Friedman:1981}, diamonds \cite{Drischler:2014} as well as our chiral results from Fig.~\ref{fig:emr_vs_density}. 
                    Most other many-body results start at unity, then tend to rise above one for low densities, and finally decrease for high densities. We find no rise in our extractions.
                    \label{fig:other_emrs}}
                \end{figure}
                       
 Comparing the chiral NN-only results from the upper panel to the chiral NN+NNN results in the lower panel, we see that
 three-neutron forces tend to slightly increase the effective mass at higher density. Similarly,                 
                when comparing the chiral and phenomenological potentials including two- and three- body forces, 
they qualitatively agree, exhibiting a comparable drop in the effective-mass ratio as the density is increased. 
The chiral interactions lead to a slightly higher effective mass at the highest densities;
prompted by this, we have carried out further calculations, with a 3N cutoff of 1.2 fm (not shown),
this time finding that this ``bending up'' is not so prominent.
From this, we conclude that the effective mass trend is relatively robust to the details of the interactions, despite having an impact on the specific values at high density. Thus, in trusting the accuracy of our QMC, and as we have taken FSE in account, these results correspond a model-independent extraction of the full effective mass ratio.

In Fig.~\ref{fig:other_emrs} we compare our chiral results using AFDMC to effective mass ratios from several many-body approaches, including both phenomenology
and \textit{ab initio} methods.                    Compared to the Skyrme energy density functional effective masses
(top panel), which have a large range of behaviours, we find partial similarity only to SIII. Others like SkP describe effective masses greater than one. The lower panel shows \textit{ab initio} many-body method values: the effective mass ratio
starts at a ratio of one, then often rises for low densities, then decreases in some fashion as the density increases. 
In 
contradistinction to this, we find an immediate, steady decrease from one; the nearest match to our trend in the literature
is with the results in Ref.~\cite{Wambach:1993}.

\section{Conclusion}
    
    In conclusion, after making some big-picture comments and summarizing earlier work on the static response of neutron matter,
    we employed periodic boundary conditions in finite particle number QMC calculations to perform a model- independent extraction of the effective mass in neutron matter as a function of density. We made use of an extrapolation prescription that used the non-interacting system to better approximate the total interacting energy at the thermodynamic limit. This allowed us to better understand the $N$ dependence, which we carefully analysed to determine an optimal particle number for our calculations. Additionally, we performed this extraction for both phenomenological and chiral two- and three- body potentials and found a general trend where the effective mass starts at unity for low density and then steadily decreases. Finally, we compared this to effective-mass ratios published in the literature, namely Skyrme energy density functionals and \textit{ab initio} many-body approaches, as well as provided some insight about the trend exhibited in our findings.
    
\

The authors wish to acknowledge the Editors of EPJ A for putting together the present Topical Issue.     
We are grateful to A. Boulet and D. Lacroix 
for many helpful conversations as well as for sharing the results from the literature which we show
in Fig.~\ref{fig:other_emrs}.
This work was supported by the Natural Sciences and Engineering Research Council (NSERC) of Canada, the
Canada Foundation for Innovation (CFI), and the Early
Researcher Award (ERA) program of the Ontario Ministry of Research, Innovation and Science. Computational resources were provided by SHARCNET and NERSC.

%
%
%

\end{document}